# The Dynamical Properties Derived from the More Generalized Lagrangian Density for A Gravitational System

**Fang-Pei Chen**

Department of Physics, Dalian University of Technology, Dalian 116024, China.

E-mail: chenfap@dlut.edu.cn

**Abstract** The studies of the generalized Einstein's Lagrangian densities without torsion are extended to those of the more generalized Lagrangian densities with torsion. The properties of the more generalized Lagrangian densities are studied systematically and thoroughly. The dynamical laws of a gravitational system such as the gravitational field equations, the conservation laws of the energy-momentum tensor densities, the conservation laws of the spin densities and the equations of motion for test particle are all derived from the Lagrangian densities. The differences between the more generalized Lagrangian densities and the generalized Einstein's Lagrangian densities are discussed.

## 1. Introduction

In the Ref. [1] the generalized Einstein's Lagrangian densities for a gravitational system

$$\sqrt{-g(x)}\, L_M(x) = \sqrt{-g(x)}\, L_M[\psi(x); \psi_{,\lambda}(x); h^i_{\cdot\mu}(x); h^i_{\cdot\mu,\lambda}(x)] \qquad (1)$$

$$\sqrt{-g(x)}\, L_G(x) = \sqrt{-g(x)}\, L_G[h^i_{\cdot\mu}(x); h^i_{\cdot\mu,\lambda}(x); h^i_{\cdot\mu,\lambda\sigma}(x)] \qquad (2)$$

have been introduced, and the field equations and conservation laws derived from them have been discussed. It has been shown in Ref. [1] that the fundamental properties of the generalized Einstein's Lagrangian densities have important implications to cosmology. In order to go a step further to understand the specific properties of the generalized Einstein's Lagrangian densities Eqs.(1,2), and for the sake of establishing a more complete theory, we will extend Eqs.(1,2) to the following expressions which are similar to those studied in Ref. [2]:

$$\sqrt{-g(x)} L_M(x) = \sqrt{-g(x)} L_M [\psi(x); \psi_{,\lambda}(x); h^i_{.\mu}(x); h^i_{.\mu,\lambda}(x); \Gamma^{ij}_{..\mu}(x)] \tag{3}$$

$$\sqrt{-g(x)} L_G(x) = \sqrt{-g(x)} L_G [h^i_{.\mu}(x); h^i_{.\mu,\lambda}(x); \Gamma^{ij}_{..\mu}(x); \Gamma^{ij}_{..\mu,\lambda}(x)] \tag{4}$$

we will call $\sqrt{-g(x)} L_M(x)$ and $\sqrt{-(g)} L_G(x)$ denoted by Eq.(3) and Eq.(4) the more generalized Lagrangian densities for a gravitational system. Where $\psi(x), h^i_{.\mu}(x), \Gamma^{ij}_{..\mu}(x)$ are the matter field, tetrad field, and frame connection field respectively. The appearance of $\Gamma^{ij}_{..\mu}(x)$ in Eq.(3) stems from that in the gauge theory of gravitation [3], $\psi_{,\mu}(x)$ must be replaced by $\psi_{|\mu}(x)$:

$\psi_{|\mu}(x) = \psi_{,\mu}(x) + \frac{1}{2}\Gamma^{ij}_{..\mu}(x)\sigma_{ij}\psi(x)$. The appearance of $\Gamma^{ij}_{..\mu,\lambda}(x)$ in Eq.(4) stems from that $\sqrt{-(g)} L_G(x)$ is composed of curvature tensor in many gravitational theories. In the Ref.[2] the most generalized Lagrangian density of matter field

$$\sqrt{-g(x)} L_M(x) = \sqrt{-g(x)} L_M [\psi(x); \psi_{,\lambda}(x); h^i_{.\mu}(x); h^i_{.\mu,\lambda}(x); \Gamma^{ij}_{..\mu}(x); \Gamma^{ij}_{..\mu,\lambda}(x);] \tag{3a}$$

has been discussed. Because of its extraordinary complexity, we only discuss Eq.(3) in this paper. It is pointed out that although the Ref.[2] has revealed many properties of the most generalized Lagrangian density of matter field denoted by Eq.(3a), there are some errors in that paper, which has been taken into consideration in analyzing Eq.(3) and Eq.(4).

Since the great majority of the fundamental matter fields are spinors, it is necessary to use tetrad field $h^i_{.\mu}(x)$ [4]. The metric field $g_{\mu\nu}(x)$ is expressed as $g_{\mu\nu}(x) = h^i_{.\mu}(x) h^j_{.\nu}(x) \eta_{ij}$, from which we have $h_i^{\cdot\mu}(x) = \eta_{ij} g^{\mu\nu}(x) h_{\cdot\nu}^j(x)$ ; $h_{i\nu,\lambda}(x) = \dfrac{\partial}{\partial x^\lambda} h_{i\nu}(x)$ ; etc. The holonomic connection field $\Gamma_{.\nu\lambda}^{\mu}(x)$ is related to $h_{.\mu}^i(x)$ and $\Gamma_{..\mu}^{ij}(x)$ by [5]

$$\Gamma_{.\nu\lambda}^{\mu}(x) = h_i^{\cdot\mu}(x)[h_{.\nu,\lambda}^i(x) + \Gamma_{.j\lambda}^i(x) h_{.\nu}^j(x)] \tag{5}$$

where $\Gamma_{.j\lambda}^i(x) = \eta_{jk}\Gamma_{..\lambda}^{ik}(x)$. In addition, $\Gamma_{..\mu}^{ij}(x) = -\Gamma_{..\mu}^{ji}(x)$. Eq.(5) can be derived from the requirement:

$$g_{\mu\nu,\lambda} - \Gamma_{.\mu\lambda}^{\sigma} g_{\sigma\nu} - \Gamma_{.\nu\lambda}^{\sigma} g_{\sigma\mu} = 0 \tag{6}$$

This requirement guarantees that lengths and angles are preserved under parallel displacement [6]. $h_{.\mu}^i(x)$ and $\Gamma_{..\mu}^{ij}(x)$ in Eqs.(3,4) are used to denote gravitational fields. If $\Gamma_{..\mu}^{ij}(x)$ are independent field variables, the torsion must appear in the space-time; if $\Gamma_{..\mu}^{ij}(x)$ all are not independent field variables, the torsion must not appear in the space-time. The torsion tensor is defined by [7]

$$T_{.\nu\lambda}^{\mu} = \tfrac{1}{2}(\Gamma_{.\nu\lambda}^{\mu} - \Gamma_{.\lambda\nu}^{\mu}) \tag{7}$$

There exists the relation [7]:

$$\Gamma_{.\nu\lambda}^{\mu} = \{{}_{\nu\lambda}^{\mu}\} + T_{\nu\lambda}^{..\mu} - T_{\lambda.\nu}^{.\mu} + T_{.\nu\lambda}^{\mu} \tag{8}$$

where 
$$\{{}_{\nu\lambda}^{\mu}\} = \tfrac{1}{2} g^{\mu\sigma}(g_{\sigma\lambda,\nu} + g_{\sigma\nu,\lambda} - g_{\nu\lambda,\sigma}) \tag{9}$$

is the Christoffel symbol. Eq.(8) can be derived from Eqs.(6,7,9). In the space-time without torsion, from Eq.(8) it is obviously $\Gamma_{.\nu\lambda}^{\mu} = \{{}_{\nu\lambda}^{\mu}\}$. In this case the relation

$$\Gamma_{..\mu}^{ij} = \tfrac{1}{2}\eta^{jk} h_k^{\cdot\nu}(h_{.\mu,\nu}^i - h_{.\nu,\mu}^i) + \tfrac{1}{2}\eta^{id} h_d^{\cdot\sigma}(h_{.\sigma,\mu}^j - h_{.\mu,\sigma}^j)$$
$$+ \tfrac{1}{2}\eta^{jk} h_k^{\cdot\nu} \eta^{id} h_d^{\cdot\sigma} \eta_{ab} h_{.\mu}^b (h_{.\sigma,\nu}^a - h_{.\nu,\sigma}^a) \tag{10}$$

can be obtained from Eqs.(5,9).

The background space-time of the Lagrangian densities Eqs.(1,2) is the space-time without torsion, but the background space-time of the Lagrangian densities Eqs.(3,4) might be the space-time with or without torsion. It can be found bellow that apart from describing the Lagrangian densities for a gravitational system with torsion, Eqs.(3,4) can be used also to describe a gravitational system without torsion. If Eqs.(3,4) are used to describe a gravitational system without torsion, it must be noted that $\Gamma^{ij}_{..\mu}(x)$ is function of $h^i_{.\mu}(x), h^i_{.\mu,\lambda}(x)$, and $\Gamma^{ij}_{..\mu,\lambda}(x)$ is function of $h^i_{.\mu}(x), h^i_{.\mu,\lambda}(x), h^i_{.\mu,\lambda\sigma}(x)$.

$\sqrt{-g(x)}\, L_M(x)$ and $\sqrt{-(g)}\, L_G(x)$ embody a great amount of information concerning many properties of a physical system including gravitation. For example, the dynamical laws such as the gravitational field equations, the conservation laws of the energy-momentum tensor densities, the conservation laws of the spin densities, the equations of motion for test particle can all be derived from $\sqrt{-g(x)}\, L_M(x)$ and $\sqrt{-(g)}\, L_G(x)$ of a gravitational system. The primary purpose of this paper is to derive the above dynamical laws from the Lagrangian densities Eqs.(3,4), and to compare the physical information contained in the Lagrangian densities Eqs.(3,4) with the physical information offered from Eqs.(1,2). We shall discuss these problems in the following sections, at this section some useful relations of differential geometry are introduced firstly.

The curvature tensor related to connection $\{^\mu_{\nu\lambda}\}$ is defined by [8]

$$\overset{(\{\})}{R^\sigma_{.\lambda\mu\nu}} = \{^\sigma_{\lambda\nu}\}_{,\mu} - \{^\sigma_{\lambda\mu}\}_{,\nu} + \{^\sigma_{\rho\mu}\}\{^\rho_{\lambda\nu}\} - \{^\sigma_{\rho\nu}\}\{^\rho_{\lambda\mu}\} \tag{11}$$

Similarly the curvature tensor related to connection $\Gamma^\mu_{.\nu\lambda}$ is defined by

$$\overset{(\Gamma)}{R^\sigma_{.\lambda\mu\nu}} = \Gamma^\sigma_{.\lambda\nu,\mu} - \Gamma^\sigma_{.\lambda\mu,\nu} + \Gamma^\sigma_{.\rho\mu}\Gamma^\rho_{.\lambda\nu} - \Gamma^\sigma_{.\rho\nu}\Gamma^\rho_{.\lambda\mu} \tag{12}$$

Eq.(8) suggests that, $\overset{(\Gamma)}{R^\sigma_{.\lambda\mu\nu}} \neq \overset{(\{\})}{R^\sigma_{.\lambda\mu\nu}}$ for the space-time with torsion; and $\overset{(\Gamma)}{R^\sigma_{.\lambda\mu\nu}} = \overset{(\{\})}{R^\sigma_{.\lambda\mu\nu}}$ only for the space-time without torsion. We can also define the curvature tensor related to the frame connection $\Gamma^{ij}_{..\mu}$ [3]:

$$R^{ij}_{..\mu\nu} = \Gamma^{ij}_{..\nu,\mu} - \Gamma^{ij}_{..\mu,\nu} + \Gamma^i_{.k\mu}\Gamma^{kj}_{..\nu} - \Gamma^i_{.k\nu}\Gamma^{kj}_{..\mu} \tag{13}$$

Using Eq.(5) it can be verified that $R^{ij}_{..\mu\nu} = \eta^{jk} h^i_{.\sigma} h^\lambda_k \overset{(\Gamma)}{R^\sigma_{.\lambda\mu\nu}}$. Using Eq.(5), the following relation for torsion tensor can also be verified:

$$T^i_{.\mu\nu}(x) = h^i_{.\lambda} T^{\lambda}_{.\mu\nu} = \frac{1}{2}(h^i_{.\mu,\nu} - h^i_{.\nu,\mu}) + \frac{1}{2}(\Gamma^i_{.j\nu} h^j_{.\mu} - \Gamma^i_{.j\mu} h^j_{.\nu}) \qquad (14)$$

## 2. Equations of fields for a gravitational system

The equations of fields for a gravitational system can be derived from [4]

$$\delta I = \delta \int (\sqrt{-g(x)} L(x)) d^4x = 0 \qquad (15)$$

where $\sqrt{-g(x)} L(x) = \sqrt{-g(x)} L_M(x) + \sqrt{-g(x)} L_G(x)$, $\delta I$ represents the variation of action integral $I = \int \sqrt{-g(x)} L(x) d^4x$ induced by the infinitesimal coordinate variation $\delta x^\mu = \xi^\mu(x)$.

Eq.(15) can be rewritten as $\delta I = \int [\delta_0 (\sqrt{-g} L) + (\xi^\alpha L)_{,\alpha}] d^4x = 0$, using Gauss' theorem, and setting $\xi^\alpha$ equal to zero at the integration limits, we get

$$\delta I = \int \delta_0 (\sqrt{-g(x)} L(x)) d^4x = 0 \qquad (15a)$$

$\delta_0(\sqrt{-g} L)$ is the variation of $\sqrt{-g} L$ corresponding to the variations of the dynamical field variable for the gravitational system at a fixed value of $x$. Eq.(15a) is equivalent to

$\delta_0 I = \int \delta_0 (\sqrt{-g(x)} L(x)) d^4x = 0$ in form.

For the gravitation with torsion, in the more generalized Lagrangian densities Eqs.(3,4), $\psi(x), h^i_{.\mu}(x), \Gamma^{ij}_{..\mu}(x)$ all are the independent dynamical field variables, then from these Lagrangian densities

$$\delta_0(\sqrt{-g}L) = \frac{\partial(\sqrt{-g}L_M)}{\partial\psi}\delta_0\psi + \frac{\partial(\sqrt{-g}L_M)}{\partial\psi_{,\lambda}}\delta_0\psi_{,\lambda} + \frac{\partial(\sqrt{-g}L)}{\partial h^i_{.\mu}}\delta_0 h^i_{.\mu}$$

$$+ \frac{\partial(\sqrt{-g}L)}{\partial h^i_{.\mu,\lambda}}\delta_0 h^i_{.\mu,\lambda} + \frac{\partial(\sqrt{-g}L)}{\partial \Gamma^{ij}_{..\mu}}\delta_0\Gamma^{ij}_{..\mu} + \frac{\partial(\sqrt{-g}L_G)}{\partial \Gamma^{ij}_{..\mu,\lambda}}\delta_0\Gamma^{ij}_{..\mu,\lambda}$$

$$= \left(\frac{\partial(\sqrt{-g}L_M)}{\partial\psi} - \frac{\partial}{\partial x^\lambda}\frac{\partial(\sqrt{-g}L_M)}{\partial\psi_{,\lambda}}\right)\delta_0\psi + \left(\frac{\partial(\sqrt{-g}L)}{\partial h^i_{.\mu}} - \frac{\partial}{\partial x^\lambda}\frac{\partial(\sqrt{-g}L)}{\partial h^i_{.\mu,\lambda}}\right)\delta_0 h^i_{.\mu}$$

$$+ \left(\frac{\partial(\sqrt{-g}L_M)}{\partial \Gamma^{ij}_{..\mu}} + \frac{\partial(\sqrt{-g}L_G)}{\partial \Gamma^{ij}_{..\mu}} - \frac{\partial}{\partial x^\lambda}\frac{\partial(\sqrt{-g}L_G)}{\partial \Gamma^{ij}_{..\mu,\lambda}}\right)\delta_0\Gamma^{ij}_{..\mu}$$

$$+ \frac{\partial}{\partial x^\lambda}\left(\frac{\partial(\sqrt{-g}L_M)}{\partial\psi_{,\lambda}}\delta_0\psi + \frac{\partial(\sqrt{-g}L)}{\partial h^i_{.\mu,\lambda}}\delta_0 h^i_{.\mu} + \frac{\partial(\sqrt{-g}L_G)}{\partial \Gamma^{ij}_{..\mu,\lambda}}\delta_0\Gamma^{ij}_{..\mu}\right) \quad (16)$$

where $\delta_0\psi(x), \delta_0 h^i_{.\mu}(x), \delta_0\Gamma^{ij}_{..\mu}(x)$ are arbitrary and independent variations, they may be or may not be symmetrical variations. Substituting Eq.(16) into Eq.(15a), using Gauss' theorem, and setting $\delta_0\psi(x), \delta_0 h^i_{.\mu}(x), \delta_0\Gamma^{ij}_{..\mu}(x)$ and their derivatives all equal to zero at the integration limits, we find

$$\left(\frac{\partial(\sqrt{-g}L_M)}{\partial\psi} - \frac{\partial}{\partial x^\lambda}\frac{\partial(\sqrt{-g}L_M)}{\partial\psi_{,\lambda}}\right)\delta_0\psi$$

$$+ \left(\frac{\partial(\sqrt{-g}L)}{\partial h^i_{.\mu}} - \frac{\partial}{\partial x^\lambda}\frac{\partial(\sqrt{-g}L)}{\partial h^i_{.\mu,\lambda}} + \right)\delta_0 h^i_{.\mu} \quad (17)$$

$$+ \left(\frac{\partial(\sqrt{-g}L_M)}{\partial \Gamma^{ij}_{..\mu}} + \frac{\partial(\sqrt{-g}L_G)}{\partial \Gamma^{ij}_{..\mu}} - \frac{\partial}{\partial x^\lambda}\frac{\partial(\sqrt{-g}L_G)}{\partial \Gamma^{ij}_{..\mu,\lambda}} + \right)\delta_0\Gamma^{ij}_{..\mu} = 0$$

Since $\psi(x), h^i_{.\mu}(x), \Gamma^{ij}_{..\mu}(x)$ are independent dynamical field variables, Eq. (17) is equivalent to the following three equations:

$$\frac{\partial(\sqrt{-g}L_M)}{\partial\psi} - \frac{\partial}{\partial x^\mu}\frac{\partial(\sqrt{-g}L_M)}{\partial\psi_{,\mu}} = 0 \quad (18)$$

$$\frac{\partial(\sqrt{-g}\,L_G)}{\partial h^i_{.\mu}} - \frac{\partial}{\partial x^\lambda}\frac{\partial(\sqrt{-g}\,L_G)}{\partial h^i_{.\mu,\lambda}} = -\frac{\partial(\sqrt{-g}\,L_M)}{\partial h^i_{.\mu}} + \frac{\partial}{\partial x^\lambda}\frac{\partial(\sqrt{-g}\,L_M)}{\partial h^i_{.\mu,\lambda}} \qquad (19)$$

$$\frac{\partial(\sqrt{-g}\,L_G)}{\partial \Gamma^{ij}_{..\mu}} - \frac{\partial}{\partial x^\lambda}\frac{\partial(\sqrt{-g}\,L_G)}{\partial \Gamma^{ij}_{..\mu,\lambda}} = -\frac{\partial(\sqrt{-g}\,L_M)}{\partial \Gamma^{ij}_{..\mu}} \qquad (20)$$

Eq.(18) is the equations of field for matter field; Eq.(19) and Eq.(20) all are the equations of field for gravitational fields. Since there are two sets of gravitational field variables $h^i_{.\mu}(x)$ and $\Gamma^{ij}_{..\mu}(x)$, hence there are two sets of equations of field for gravitational fields.

In the generalized Einstein's Lagrangian densities Eqs.(1,2) which describe the gravitational system without torsion, the independent dynamical field variables are only $\psi(x), h^i_{.\mu}(x)$, their equations of field have been given as follows [1]:

$$\frac{\partial(\sqrt{-g}\,L_M)}{\partial \psi} - \frac{\partial}{\partial x^\mu}\frac{\partial(\sqrt{-g}\,L_M)}{\partial \psi_{,\mu}} = 0 \qquad (18')$$

$$\frac{\partial(\sqrt{-g}\,L_G)}{\partial h^i_{.\mu}} - \frac{\partial}{\partial x^\lambda}\frac{\partial(\sqrt{-g}\,L_G)}{\partial h^i_{.\mu,\lambda}} + \frac{\partial^2}{\partial x^\lambda \partial x^\sigma}\frac{\partial(\sqrt{-g}\,L_G)}{\partial h^i_{.\mu,\lambda\sigma}}$$
$$= -\frac{\partial(\sqrt{-g}\,L_M)}{\partial h^i_{.\mu}} + \frac{\partial}{\partial x^\lambda}\frac{\partial(\sqrt{-g}\,L_M)}{\partial h^i_{.\mu,\lambda}} \qquad (21)$$

In this case there is only one set of gravitational field variables $h^i_{.\mu}(x)$, hence there is only one set of equations of field *i.e.* Eq.(21) for gravitational fields. It must be noted that although Eqs.(1,2) are the special case of Eqs.(3,4) if using the relation Eq.(10), Eq.(21) can not be deduced directly from Eq.(19) and Eq.(20).

## 3. The symmetry of the Lagrangian densities for a gravitational system

Symmetries exist universally in physical systems, one fundamental symmetry of a gravitational system is that the action integrals

$$I_M = \int \sqrt{-g(x)}\, L_M(x)\, d^4x \qquad I_G = \int \sqrt{-g(x)}\, L_G(x)\, d^4x \qquad \text{and}$$

$$I = I_M + I_G = \int \sqrt{-g(x)}\, (L_M(x) + L_G(x))\, d^4x$$

satisfy $\delta I_M = 0$, $\delta I_G = 0$ and $\delta I = 0$ respectively under the following two simultaneous transformations [3,6]:

(1), the infinitesimal general coordinate transformation

$$x^\mu \to x'^\mu = x^\mu + \xi^\mu(x) \tag{22}$$

(2), the local Lorentz transformation of tetrad frame

$$e_i(x) \to e'_i(x') = e_i(x) - \varepsilon^{mn}(x)\, \delta_m^j\, \eta_{ni}\, e_j(x) \tag{23}$$

The symmetry (1) is precisely the symmetry of local space-time translations.

The sufficient condition of an action integral $I = \int \sqrt{-g(x)}\, L(x)\, d^4x$ being $\delta I = 0$ under above transformations is [3,9]:

$$\delta_0(\sqrt{-g}\, L) + (\xi^\mu \sqrt{-g}\, L)_{,\mu} \equiv 0 \tag{24}$$

where $\delta_0$ represent the variation at a fixed value of $x$. Evidently there are also the relations

$$\delta_0(\sqrt{-g}\, L_M) + (\xi^\mu \sqrt{-g}\, L_M)_{,\mu} \equiv 0 \;;\; \delta_0(\sqrt{-g}\, L_G) + (\xi^\mu \sqrt{-g}\, L_G)_{,\mu} \equiv 0 \tag{25}$$

If there exists only the symmetry (2), Eqs.(24,25) reduce to $\delta_0(\sqrt{-g}\, L) \equiv 0$, $\delta_0(\sqrt{-g}\, L_M) \equiv 0$ and $\delta_0(\sqrt{-g}\, L_G) \equiv 0$ respectively.

For the more generalized Lagrangian densities Eqs.(3,4), we have

$$\delta_0(\sqrt{-g}\, L_M) = \frac{\partial(\sqrt{-g}\, L_M)}{\partial \psi} \delta_0 \psi + \frac{\partial(\sqrt{-g}\, L_M)}{\partial \psi_{,\lambda}} \delta_0 \psi_{,\lambda} + \frac{\partial(\sqrt{-g}\, L_M)}{\partial h^i_{.\mu}} \delta_0 h^i_{.\mu}$$
$$+ \frac{\partial(\sqrt{-g}\, L_M)}{\partial h^i_{.\mu,\lambda}} \delta_0 h^i_{.\mu,\lambda} + \frac{\partial(\sqrt{-g}\, L_M)}{\partial \Gamma^{ij}_{..\mu}} \delta_0 \Gamma^{ij}_{..\mu} \tag{26}$$

$$\delta_0(\sqrt{-g}\,L_G) = \frac{\partial(\sqrt{-g}\,L_G)}{\partial h^i_{.\mu}}\delta_0 h^i_{.\mu} + \frac{\partial(\sqrt{-g}\,L_G)}{\partial h^i_{.\mu,\lambda}}\delta_0 h^i_{.\mu,\lambda}$$
$$+ \frac{\partial(\sqrt{-g}\,L_G)}{\partial \Gamma^{ij}_{..\mu}}\delta_0 \Gamma^{ij}_{..\mu} + \frac{\partial(\sqrt{-g}\,L_G)}{\partial \Gamma^{ij}_{..\mu,\lambda}}\delta_0 \Gamma^{ij}_{..\mu,\lambda} \quad (27)$$

As $\psi(x)$ is spinor and $h^i_{.\mu}(x)$ is both tetrad Lorentz vector and coordinate vector, and $\Gamma^{ij}_{..\mu}(x)$ transforms as

$$\Gamma^{ij}_{..\mu}(x) \to \Gamma'^{ij}_{..\mu}(x) = \frac{\partial x^{\nu}}{\partial x'^{\mu}}\Lambda^i_a(x)\Lambda^j_b(x)\Gamma^{ab}_{..\nu}(x) - \eta^{ak}\Lambda^j_k(x)\frac{\partial}{\partial x'^{\mu}}\Lambda^i_a(x)$$

under the infinitesimal general coordinate transformation and the local Lorentz transformation of tetrad frame, it is not difficult to derive the following induced variations [5]:

$$\delta_0 \psi(x) = \frac{1}{2}\varepsilon^{mn}(x)\sigma_{mn}\psi(x) - \xi^{\alpha}(x)\psi_{,\alpha}(x) \quad (28)$$

$$\delta_0 \psi_{,\lambda}(x) = \frac{1}{2}\varepsilon^{mn}(x)\sigma_{mn}\psi_{,\lambda}(x) + \frac{1}{2}\varepsilon^{mn}_{,\lambda}(x)\sigma_{mn}\psi(x) - \xi^{\alpha}(x)\psi_{,\alpha\lambda}(x)$$
$$-\xi^{\alpha}_{,\lambda}(x)\psi_{,\alpha}(x) \quad (29)$$

$$\delta_0 h^i_{.\mu}(x) = \varepsilon^{mn}(x)\delta^i_m \eta_{nj} h^j_{.\mu}(x) - \xi^{\alpha}_{,\mu}(x)h^i_{.\alpha}(x) - \xi^{\alpha}(x)h^i_{.\mu,\alpha}(x) \quad (30)$$

$$\delta_0 h^i_{.\mu,\lambda}(x) = \varepsilon^{mn}(x)\delta^i_m \eta_{nj} h^j_{.\mu,\lambda}(x) + \varepsilon^{mn}_{,\lambda}(x)\delta^i_m \eta_{nj} h^j_{.\mu}(x) - \xi^{\alpha}_{,\mu}(x)h^i_{.\alpha,\lambda}(x)$$
$$-\xi^{\alpha}_{,\mu\lambda}(x)h^i_{.\alpha}(x) - \xi^{\alpha}(x)h^i_{.\mu,\alpha\lambda}(x) - \xi^{\alpha}_{,\lambda}(x)h^i_{.\mu,\alpha} \quad (31)$$

$$\delta_0 \Gamma^{ij}_{..\mu}(x) = \varepsilon^{mn}(x)(\delta^i_m \eta_{nk}\Gamma^{kj}_{..\mu}(x) + \delta^j_m \eta_{nk}\Gamma^{ik}_{..\mu}(x)) - \xi^{\alpha}_{,\mu}(x)\Gamma^{ij}_{..\alpha}(x)$$
$$-\varepsilon^{ij}_{,\mu}(x) - \xi^{\alpha}(x)\Gamma^{ij}_{..\mu,\alpha}(x) \quad (32)$$

$$\delta_0\Gamma^{ij}_{..\mu,\lambda} = \varepsilon^{mn}(x)(\delta^i_m\eta_{nk}\Gamma^{kj}_{..\mu,\lambda}(x)+\delta^j_m\eta_{nk}\Gamma^{ik}_{..\mu,\lambda}(x))+\varepsilon^{mn}_{,\lambda}(x)(\delta^i_m\eta_{nk}\Gamma^{kj}_{..\mu}(x)$$

$$+\delta^j_m\eta_{nk}\Gamma^{ik}_{..\mu}(x))-\xi^\alpha_{,\mu}(x)\Gamma^{ij}_{..\alpha,\lambda}(x)-\xi^\alpha_{,\mu\lambda}(x)\Gamma^{ij}_{..\alpha}(x)-\varepsilon^{ij}_{,\mu\lambda}(x)-\xi^\alpha(x)\Gamma^{ij}_{..\mu,\alpha\lambda}(x)$$

$$\xi^\alpha_{,\lambda}(x)\Gamma^{ij}_{..\mu,\alpha}(x)$$

(33)

Substituting Eqs.(28 -33) into Eq.(26) and Eq.(27); using $\delta_0(\sqrt{-g}\Lambda)+(\xi^\mu\sqrt{-g}\Lambda)_{,\mu}\equiv 0$, where $\Lambda=L_M$ or $\Lambda=L_G$ or $\Lambda=L_M+L_G$. Because of the independent arbitrariness of $\varepsilon^{mn}_{,}(x)$, $\varepsilon^{mn}_{,\lambda}(x)$, $\varepsilon^{mn}_{,\lambda\sigma}(x)$, $\xi^\alpha(x)$, $\xi^\alpha_{,\mu}(x)$ and $\xi^\alpha_{,\mu\lambda}(x)$, we obtain the following identities ( if $\Lambda=L_G$, $\frac{\partial(\sqrt{-g}\Lambda)}{\partial\psi}=0$, $\frac{\partial(\sqrt{-g}\Lambda)}{\partial\psi_{,\lambda}}=0$; if $\Lambda=L_M$, $\frac{\partial(\sqrt{-g}\Lambda)}{\partial\Gamma^{ij}_{..\mu,\lambda}}=0$ ):

$$\frac{1}{2}\frac{\partial(\sqrt{-g}\Lambda)}{\partial\psi}\sigma_{mn}\psi+\frac{1}{2}\frac{\partial(\sqrt{-g}\Lambda)}{\partial\psi_{,\lambda}}\sigma_{mn}\psi_{,\lambda}+\frac{\partial(\sqrt{-g}\Lambda)}{\partial h^m_{.\mu}}h_{n\mu}$$

$$+\frac{\partial(\sqrt{-g}\Lambda)}{\partial h^m_{.\mu,\lambda}}h_{n\mu,\lambda}+2\frac{\partial(\sqrt{-g}\Lambda)}{\partial\Gamma^{km}_{..\mu}}\Gamma^k_{.n\mu}+2\frac{\partial(\sqrt{-g}\Lambda)}{\partial\Gamma^{km}_{..\mu,\lambda}}\Gamma^k_{.n\mu,\lambda}=0$$

(34)

$$\frac{1}{2}\frac{\partial(\sqrt{-g}\Lambda)}{\partial\psi_{,\lambda}}\sigma_{mn}\psi+\frac{\partial(\sqrt{-g}\Lambda)}{\partial h^m_{.\mu,\lambda}}h_{n\mu}+2\frac{\partial(\sqrt{-g}\Lambda)}{\partial\Gamma^{km}_{..\mu,\lambda}}\Gamma^k_{.n\mu}=\frac{\partial(\sqrt{-g}\Lambda)}{\partial\Gamma^{mn}_{..\lambda}} \qquad (35)$$

$$\frac{\partial(\sqrt{-g}\Lambda)}{\partial\Gamma^{mn}_{..\mu,\nu}}=-\frac{\partial(\sqrt{-g}\Lambda)}{\partial\Gamma^{mn}_{..\nu,\mu}} \qquad (36)$$

$$\frac{\partial(\sqrt{-g}\,\Lambda)}{\partial\psi}\psi_{,\alpha}+\frac{\partial(\sqrt{-g}\,\Lambda)}{\partial\psi_{,\lambda}}\psi_{,\lambda\alpha}+\frac{\partial(\sqrt{-g}\,\Lambda)}{\partial h^{i}_{.\mu}}h^{i}_{.\mu,\alpha}$$

$$+\frac{\partial(\sqrt{-g}\,\Lambda)}{\partial h^{i}_{.\mu,\lambda}}h^{i}_{.\mu,\alpha\lambda}+\frac{\partial(\sqrt{-g}\,\Lambda)}{\partial\Gamma^{ij}_{..\mu}}\Gamma^{ij}_{..\mu,\alpha}+\frac{\partial(\sqrt{-g}\,\Lambda)}{\partial\Gamma^{ij}_{..\mu,\lambda}}\Gamma^{ij}_{..\mu,\alpha\lambda}-(\sqrt{-g}\,\Lambda)_{,\alpha}=0 \quad (37)$$

$$\frac{\partial(\sqrt{-g}\,\Lambda)}{\partial\psi_{,\lambda}}\psi_{,\alpha}+\frac{\partial(\sqrt{-g}\,\Lambda)}{\partial h^{i}_{.\lambda}}h^{i}_{\alpha}+\frac{\partial(\sqrt{-g}\,\Lambda)}{\partial h^{i}_{.\mu,\lambda}}(h^{i}_{.\mu,\alpha}-h^{i}_{.\alpha,\mu})$$

$$+\frac{\partial(\sqrt{-g}\,\Lambda)}{\partial\Gamma^{ij}_{..\lambda}}\Gamma^{ij}_{..\alpha}+\frac{\partial(\sqrt{-g}\,\Lambda)}{\partial\Gamma^{ij}_{..\mu,\lambda}}(\Gamma^{ij}_{..\mu,\alpha}-\Gamma^{ij}_{..\alpha,\mu})-\sqrt{-g}\,\Lambda\,\delta^{\lambda}_{\alpha}=0 \quad (38)$$

$$\frac{\partial(\sqrt{-g}\,\Lambda)}{\partial h^{i}_{.\mu,\lambda}}h^{i}_{.\alpha}+\frac{\partial(\sqrt{-g}\,\Lambda)}{\partial\Gamma^{ij}_{..\mu,\lambda}}\Gamma^{ij}_{..\alpha}=0 \quad (39)$$

From Eq.(39) and Eq.(36), it is found that there must exist another identity:

$$\frac{\partial(\sqrt{-g}\,\Lambda)}{\partial h^{i}_{.\mu,\nu}}=-\frac{\partial(\sqrt{-g}\,\Lambda)}{\partial h^{i}_{.\nu,\mu}} \quad (40)$$

For the generalized Einstein's Lagrangian densities Eqs.(1,2) we have

$$\delta_{0}(\sqrt{-g}\,L_{M})=\frac{\partial(\sqrt{-g}\,L_{M})}{\partial\psi}\delta_{0}\psi+\frac{\partial(\sqrt{-g}\,L_{M})}{\partial\psi_{,\lambda}}\delta_{0}\psi_{,\lambda}$$

$$+\frac{\partial(\sqrt{-g}\,L_{M})}{\partial h^{i}_{.\mu}}\delta_{0}h^{i}_{.\mu}+\frac{\partial(\sqrt{-g}\,L_{M})}{\partial h^{i}_{.\mu,\lambda}}\delta_{0}h^{i}_{.\mu,\lambda} \quad (41)$$

$$\delta_{0}(\sqrt{-g}\,L_{G})=\frac{\partial(\sqrt{-g}\,L_{G})}{\partial h^{i}_{.\mu}}\delta_{0}h^{i}_{.\mu}+\frac{\partial(\sqrt{-g}\,L_{G})}{\partial h^{i}_{.\mu,\lambda}}\delta_{0}h^{i}_{.\mu,\lambda}+\frac{\partial(\sqrt{-g}\,L_{G})}{\partial h^{i}_{.\mu,\lambda\sigma}}\delta_{0}h^{i}_{.\mu,\lambda\sigma} \quad (42)$$

Substituting Eqs.(28 -31) and the induced variation

$$\delta_0 h^i_{.\mu,\lambda\sigma}(x) = \varepsilon^{mn}(x)\delta^i_m \eta_{nj} h^j_{.\mu,\lambda\sigma}(x) + \varepsilon^{mn}_{,\sigma}(x)\delta^i_m \eta_{nj} h^j_{.\mu,\lambda}(x)$$

$$+ \varepsilon^{mn}_{,\lambda}(x)\delta^i_m \eta_{nj} h^j_{.\mu,\sigma}(x) + \varepsilon^{mn}_{,\lambda\sigma}(x)\delta^i_m \eta_{nj} h^j_{.\mu}(x) - \xi^\alpha_{,\mu}(x) h^i_{.\alpha,\lambda\sigma}(x)$$

$$- \xi^\alpha_{,\mu\sigma}(x) h^i_{.\alpha,\lambda}(x) - \xi^\alpha_{,\mu\lambda}(x) h^i_{.\alpha,\sigma}(x) - \xi^\alpha_{,\mu\lambda\sigma}(x) h^i_{.\alpha}(x) - \xi^\alpha(x) h^i_{.\mu,\alpha\lambda\sigma}(x)$$

$$- \xi^\alpha_{,\sigma}(x) h^i_{.\mu,\alpha\lambda}(x) - \xi^\alpha_{,\lambda}(x) h^i_{.\mu,\alpha\sigma}(x) - \xi^\alpha_{,\lambda\sigma}(x) h^i_{.\mu,\alpha}(x)$$
(43)

into Eq.(41) and Eq.(42); using $\delta_0(\sqrt{-g}\Lambda) + (\xi^\mu \sqrt{-g}\Lambda)_{,\mu} \equiv 0$, where $\Lambda = L_M$ or $\Lambda = L_G$ or $\Lambda = L_M + L_G$; owing to the independent arbitrariness of $\varepsilon^{mn}(x)$, $\varepsilon^{mn}_{,\lambda}(x)$, $\varepsilon^{mn}_{,\lambda\sigma}(x)$, $\xi^\alpha(x)$, $\xi^\alpha_{,\mu}(x)$, $\xi^\alpha_{,\mu\lambda}(x)$ and $\xi^\alpha_{,\mu\lambda\sigma}(x)$, we obtain another set of identities [1] ( if $\Lambda = L_G$, $\frac{\partial(\sqrt{-g}\Lambda)}{\partial \psi} = 0$, $\frac{\partial(\sqrt{-g}\Lambda)}{\partial \psi_{,\lambda}} = 0$; if $\Lambda = L_M$, $\frac{\partial(\sqrt{-g}\Lambda)}{\partial h^i_{.\mu,\lambda\sigma}} = 0$ ):

$$\frac{1}{2}\frac{\partial(\sqrt{-g}\Lambda)}{\partial \psi}\sigma_{mn}\psi + \frac{1}{2}\frac{\partial(\sqrt{-g}\Lambda)}{\partial \psi_{,\lambda}}\sigma_{mn}\psi_{,\lambda} + \frac{\partial(\sqrt{-g}\Lambda)}{\partial h^m_{.\mu}} h_{n\mu}$$

$$+ \frac{\partial(\sqrt{-g}\Lambda)}{\partial h^m_{.\mu,\lambda}} h_{n\mu,\lambda} + \frac{\partial(\sqrt{-g}\Lambda)}{\partial h^m_{.\mu,\lambda\sigma}} h_{n\mu,\lambda\sigma} = 0$$
(44)

$$\frac{1}{2}\frac{\partial(\sqrt{-g}\Lambda)}{\partial \psi_{,\lambda}}\sigma_{mn}\psi + \frac{\partial(\sqrt{-g}\Lambda)}{\partial h^m_{.\mu,\lambda}} h_{n\mu} + 2\frac{\partial(\sqrt{-g}\Lambda)}{\partial h^m_{.\mu,\lambda\sigma}} h_{n\mu,\sigma} = 0$$
(45)

$$\frac{\partial(\sqrt{-g}\Lambda)}{\partial h^m_{.\mu,\lambda\sigma}} h_{n\mu} = \frac{\partial(\sqrt{-g}\Lambda)}{\partial h^n_{.\mu,\lambda\sigma}} h_{m\mu}$$
(46)

$$\frac{\partial(\sqrt{-g}\Lambda)}{\partial \psi}\psi_{,\alpha}+\frac{\partial(\sqrt{-g}\Lambda)}{\partial \psi_{,\lambda}}\psi_{,\lambda\alpha}+\frac{\partial(\sqrt{-g}\Lambda)}{\partial h^i_{.\mu}}h^i_{.\mu,\alpha}$$
$$+\frac{\partial(\sqrt{-g}\Lambda)}{\partial h^i_{.\mu,\lambda}}h^i_{.\mu,\lambda\alpha}+\frac{\partial(\sqrt{-g}\Lambda)}{\partial h^i_{.\mu,\lambda\sigma}}h^i_{.\mu,\lambda\sigma\alpha}-(\sqrt{-g}\Lambda)_{,\alpha}=0 \qquad (47)$$

$$\frac{\partial(\sqrt{-g}\Lambda)}{\partial \psi_{,\lambda}}\psi_{,\alpha}+\frac{\partial(\sqrt{-g}\Lambda)}{\partial h^i_{.\lambda}}h^i_{.\alpha}+\frac{\partial(\sqrt{-g}\Lambda)}{\partial h^i_{.\mu,\lambda}}h^i_{.\mu,\alpha}+\frac{\partial(\sqrt{-g}\Lambda)}{\partial h^i_{.\lambda,\mu}}h^i_{.\alpha,\mu}$$
$$+\frac{\partial(\sqrt{-g}\Lambda)}{\partial h^i_{.\lambda,\mu\sigma}}h^i_{.\alpha,\mu\sigma}+2\frac{\partial(\sqrt{-g}\Lambda)}{\partial h^i_{.\mu,\lambda\sigma}}h^i_{.\mu,\sigma\alpha}-\sqrt{-g}\Lambda\,\delta^\lambda_\alpha=0 \qquad (48)$$

$$\frac{\partial(\sqrt{-g}\Lambda)}{\partial h^i_{.\mu,\lambda}}h^i_{.\alpha}+\frac{\partial(\sqrt{-g}\Lambda)}{\partial h^i_{.\mu,\lambda\sigma}}h^i_{.\alpha,\sigma}+\frac{\partial(\sqrt{-g}\Lambda)}{\partial h^i_{.\sigma,\lambda\mu}}h^i_{.\sigma,\alpha}-\frac{\partial}{\partial x^\sigma}(\frac{\partial(\sqrt{-g}\Lambda)}{\partial h^i_{.\mu,\lambda\sigma}})h^i_{.\alpha}$$
$$=-\frac{\partial}{\partial x^\sigma}(\frac{\partial(\sqrt{-g}\Lambda)}{\partial h^i_{.\mu,\lambda\sigma}}h^i_{.\alpha}) \qquad (49)$$

$$\frac{\partial(\sqrt{-g}\Lambda)}{\partial h^i_{.\mu,\lambda\sigma}}h^i_{.\alpha}+\frac{\partial(\sqrt{-g}\Lambda)}{\partial h^i_{.\lambda,\sigma\mu}}h^i_{.\alpha}+\frac{\partial(\sqrt{-g}\Lambda)}{\partial h^i_{.\sigma,\mu\lambda}}h^i_{.\alpha}=0 \qquad (50)$$

From Eq.(50) another identity:

$$\frac{\partial^3}{\partial x^\mu \partial x^\lambda \partial x^\sigma}(\frac{\partial(\sqrt{-g}\Lambda)}{\partial h^i_{.\mu,\lambda\sigma}}h^i_{.\alpha})=0 \qquad (51)$$

can be deduced.

Eqs.(1,2) are used only to describe a gravitational system without torsion; when Eqs.(3,4) are used only to describe a gravitational system with torsion, there is not any relation between the above two sets of identities Eqs.(34 -39) and Eqs.(44-50). If Eqs.(3,4) are used to describe a gravitational system without torsion, there must be some relations between Eqs.(34 -39) and Eqs.(44-50), now we discuss these relations.

When Eqs.(3,4) are used also to describe a gravitational system without torsion, from Eq.(10) they must be expressed by

$$\sqrt{-g(x)}\,L_M(x) = \sqrt{-g(x)}\,L_M[\psi(x);\psi_{,\lambda}(x);h^i_{.\mu}(x);h^i_{.\mu,\lambda}(x);$$
$$\Gamma^{ab}_{..\alpha}[h^i_{.\mu}(x);h^i_{.\mu,\lambda}(x)]] = \sqrt{-g(x)}\,L^*_M[\psi(x);\psi_{,\lambda}(x);h^i_{.\mu}(x);h^i_{.\mu,\lambda}(x)] \quad (52)$$

$$\sqrt{-g(x)}\,L_G(x) = \sqrt{-g(x)}\,L_G[h^i_{.\mu}(x);h^i_{.\mu,\lambda}(x);$$
$$\Gamma^{ab}_{..\alpha}[h^i_{.\mu}(x);h^i_{.\mu,\lambda}(x)];\Gamma^{ab}_{..\alpha,\beta}[h^i_{.\mu}(x);h^i_{.\mu,\lambda}(x);h^i_{.\mu,\lambda\sigma}]] \quad (53)$$
$$= \sqrt{-g(x)}\,L^*_G[h^i_{.\mu}(x);h^i_{.\mu,\lambda}(x);h^i_{.\mu,\lambda\sigma}(x)]$$

Therefore from Eq.(52)

$$\delta_0(\sqrt{-g}\,L_M) = \frac{\partial(\sqrt{-g}\,L_M)}{\partial\psi}\delta_0\psi + \frac{\partial(\sqrt{-g}\,L_M)}{\partial\psi_{,\lambda}}\delta_0\psi_{,\lambda} + \left(\frac{\partial(\sqrt{-g}\,L_M)}{\partial h^i_{.\mu}}\right)_\Gamma \delta_0 h^i_{.\mu}$$
$$+ \left(\frac{\partial(\sqrt{-g}\,L_M)}{\partial h^i_{.\mu,\lambda}}\right)_\Gamma \delta_0 h^i_{.\mu,\lambda} + \frac{\partial(\sqrt{-g}\,L_M)}{\partial\Gamma^{ab}_{..\alpha}}\frac{\partial\Gamma^{ab}_{..\alpha}}{\partial h^i_{.\mu}}\delta_0 h^i_{.\mu} + \frac{\partial(\sqrt{-g}\,L_M)}{\partial\Gamma^{ab}_{..\alpha}}\frac{\partial\Gamma^{ab}_{..\alpha}}{\partial h^i_{.\mu,\lambda}}\delta_0 h^i_{.\mu,\lambda} \quad (54)$$
$$= \delta_0(\sqrt{-g}\,L^*_M) = \frac{\partial(\sqrt{-g}\,L^*_M)}{\partial\psi}\delta_0\psi + \frac{\partial(\sqrt{-g}\,L^*_M)}{\partial\psi_{,\lambda}}\delta_0\psi_{,\lambda} + \frac{\partial(\sqrt{-g}\,L^*_M)}{\partial h^i_{.\mu}}\delta_0 h^i_{.\mu}$$
$$+ \frac{\partial(\sqrt{-g}\,L^*_M)}{\partial h^i_{.\mu,\lambda}}\delta_0 h^i_{.\mu,\lambda}$$

Where $\left(\frac{\partial}{\partial}\right)_\Gamma$ denote the partial derivative at constant $\Gamma^{ij}_{..\mu}(x)$. Hence we have

$$\frac{\partial(\sqrt{-g}\,L^*_M)}{\partial\psi} = \frac{\partial(\sqrt{-g}\,L_M)}{\partial\psi}, \quad \frac{\partial(\sqrt{-g}\,L^*_M)}{\partial\psi_{,\lambda}} = \frac{\partial(\sqrt{-g}\,L_M)}{\partial\psi_{,\lambda}} \quad (55)$$

$$\frac{\partial(\sqrt{-g}\,L^*_M)}{\partial h^i_{.\mu}} = \left(\frac{\partial(\sqrt{-g}\,L_M)}{\partial h^i_{.\mu}}\right)_\Gamma + \frac{\partial(\sqrt{-g}\,L_M)}{\partial\Gamma^{ab}_{..\alpha}}\frac{\partial\Gamma^{ab}_{..\alpha}}{\partial h^i_{.\mu}} \quad (56)$$

$$\frac{\partial(\sqrt{-g}\,L_M^*)}{\partial h^i_{.\mu,\lambda}} = \left(\frac{\partial(\sqrt{-g}\,L_M)}{\partial h^i_{.\mu,\lambda}}\right)_\Gamma + \frac{\partial(\sqrt{-g}\,L_M)}{\partial \Gamma^{ab}_{..\alpha}}\frac{\partial \Gamma^{ab}_{..\alpha}}{\partial h^i_{.\mu,\lambda}} \tag{57}$$

From Eq.(53)

$$\delta_0(\sqrt{-g}\,L_G) = \left(\frac{\partial(\sqrt{-g}\,L_G)}{\partial h^i_{.\mu}}\right)_\Gamma \delta_0 h^i_{.\mu} + \left(\frac{\partial(\sqrt{-g}\,L_G)}{\partial h^i_{.\mu,\lambda}}\right)_\Gamma \delta_0 h^i_{.\mu,\lambda}$$

$$+ \frac{\partial(\sqrt{-g}\,L_G)}{\partial \Gamma^{ab}_{..\alpha}}\frac{\partial \Gamma^{ab}_{..\alpha}}{\partial h^i_{.\mu}}\delta_0 h^i_{.\mu} + \frac{\partial(\sqrt{-g}\,L_G)}{\partial \Gamma^{ab}_{..\alpha}}\frac{\partial \Gamma^{ab}_{..\alpha}}{\partial h^i_{.\mu,\lambda}}\delta_0 h^i_{.\mu,\lambda}$$

$$+ \frac{\partial(\sqrt{-g}\,L_G)}{\partial \Gamma^{ab}_{..\alpha,\beta}}\frac{\partial \Gamma^{ab}_{..\alpha,\beta}}{\partial h^i_{.\mu}}\delta_0 h^i_{.\mu} + \frac{\partial(\sqrt{-g}\,L_G)}{\partial \Gamma^{ab}_{..\alpha,\beta}}\frac{\partial \Gamma^{ab}_{..\alpha,\beta}}{\partial h^i_{.\mu,\lambda}}\delta_0 h^i_{.\mu,\lambda} + \frac{\partial(\sqrt{-g}\,L_G)}{\partial \Gamma^{ab}_{..\alpha,\beta}}\frac{\partial \Gamma^{ab}_{..\alpha,\beta}}{\partial h^i_{.\mu,\lambda\sigma}}\delta_0 h^i_{.\mu,\lambda\sigma}$$

$$= \delta_0(\sqrt{-g}\,L_G^*) = \frac{\partial(\sqrt{-g}\,L_G^*)}{\partial h^i_{.\mu}}\delta_0 h^i_{.\mu} + \frac{\partial(\sqrt{-g}\,L_G^*)}{\partial h^i_{.\mu,\lambda}}\delta_0 h^i_{.\mu,\lambda} + \frac{\partial(\sqrt{-g}\,L_G^*)}{\partial h^i_{.\mu,\lambda\sigma}}\delta_0 h^i_{.\mu,\lambda\sigma}$$

$$\tag{58}$$

Hence we have

$$\frac{\partial(\sqrt{-g}\,L_G^*)}{\partial h^i_{.\mu}} = \left(\frac{\partial(\sqrt{-g}\,L_G)}{\partial h^i_{.\mu}}\right)_\Gamma + \frac{\partial(\sqrt{-g}\,L_G)}{\partial \Gamma^{ab}_{..\alpha}}\frac{\partial \Gamma^{ab}_{..\alpha}}{\partial h^i_{.\mu}} + \frac{\partial(\sqrt{-g}\,L_G)}{\partial \Gamma^{ab}_{..\alpha,\beta}}\frac{\partial \Gamma^{ab}_{..\alpha,\beta}}{\partial h^i_{.\mu}} \tag{59}$$

$$\frac{\partial(\sqrt{-g}\,L_G^*)}{\partial h^i_{.\mu,\lambda}} = \left(\frac{\partial(\sqrt{-g}\,L_G)}{\partial h^i_{.\mu,\lambda}}\right)_\Gamma + \frac{\partial(\sqrt{-g}\,L_G)}{\partial \Gamma^{ab}_{..\alpha}}\frac{\partial \Gamma^{ab}_{..\alpha}}{\partial h^i_{.\mu,\lambda}} + \frac{\partial(\sqrt{-g}\,L_G)}{\partial \Gamma^{ab}_{..\alpha,\beta}}\frac{\partial \Gamma^{ab}_{..\alpha,\beta}}{\partial h^i_{.\mu,\lambda}} \tag{60}$$

$$\frac{\partial(\sqrt{-g}\,L_G^*)}{\partial h^i_{.\mu,\lambda\sigma}} = \frac{\partial(\sqrt{-g}\,L_G)}{\partial \Gamma^{ab}_{..\alpha,\beta}}\frac{\partial \Gamma^{ab}_{..\alpha,\beta}}{\partial h^i_{.\mu,\lambda\sigma}} \tag{61}$$

On the other hand we also have:

$$\frac{\partial(\sqrt{-g}\,\Lambda)}{\partial \Gamma^{ab}_{..\alpha}}\delta_0 \Gamma^{ab}_{..\alpha} = \frac{\partial(\sqrt{-g}\,\Lambda)}{\partial \Gamma^{ab}_{..\alpha}}\left(\frac{\partial \Gamma^{ab}_{..\alpha}}{\partial h^i_{.\mu}}\delta_0 h^i_{.\mu} + \frac{\partial \Gamma^{ab}_{..\alpha}}{\partial h^i_{.\mu,\lambda}}\delta_0 h^i_{.\mu,\lambda}\right) \tag{62}$$

$$\frac{\partial(\sqrt{-g}\Lambda)}{\partial\Gamma^{ab}_{..\alpha,\beta}}\delta_0\Gamma^{ab}_{..\alpha,\beta} = \frac{\partial(\sqrt{-g}\Lambda)}{\partial\Gamma^{ab}_{..\alpha,\beta}}(\frac{\partial\Gamma^{ab}_{..\alpha,\beta}}{\partial h^i_{.\mu}}\delta_0 h^i_{.\mu} + \frac{\partial\Gamma^{ab}_{..\alpha,\beta}}{\partial h^i_{.\mu,\lambda}}\delta_0 h^i_{.\mu,\lambda}$$
$$+\frac{\partial\Gamma^{ab}_{..\alpha,\beta}}{\partial h^i_{.\mu,\lambda\sigma}}\delta_0 h^i_{.\mu,\lambda\sigma})$$
(63)

where $\Lambda = L_M$ or $\Lambda = L_G$ or $\Lambda = L_M + L_G$ ( if $\Lambda = L_M$, $\frac{\partial(\sqrt{-g}\Lambda)}{\partial\Gamma^{ab}_{..\alpha,\beta}} = 0$ ). Utilizing these relations and relations Eqs.(55-57,59-61), it can be proved that the identities Eqs.(34-39) are equivalent to the identities Eqs.(44-50).

Although Eqs.(3,4) can be used to describe a gravitational system without torsion, but due to the complex nature of their mathematical expressions for the non-torsion case, for practical purposes there is always using Eqs.(1,2) to describe a gravitational system without torsion. However, under some circumstances certain conclusions suitable for gravitational system both with and without torsion can be derived from Eqs.(3,4); an example will be given in next section.

## 4. Possible forms of gravitational Lagrangian

By the requirement that the action integrals of a gravitational system is invariant under the following two simultaneous transformations : (1), the infinitesimal general coordinate transformation

$x^\mu \to x'^\mu = x^\mu + \xi^\mu(x)$ ; (2), the local Lorentz transformation of tetrad frame

$e_i(x) \to e'_i(x') = e_i(x) - \varepsilon^{mn}(x)\delta^j_m \eta_{ni} e_j(x)$ ; one can show that the possible forms of the more

generalized gravitational Lagrangian density Eq.(4) might be:

$$\sqrt{-g}\, L_G(x) = \sqrt{-g}\, L_G^{RT}(x) = \sqrt{-g}\, L_G^{RT}[R^{ij}_{..\mu\nu}(x)\,; T^i_{.\mu\nu}(x)\,; h^i_{.\mu}(x)\,]$$
(64)

$\sqrt{-g}\, L_G(x)$ denoted by Eq.(4) must satisfy the identities Eqs.(34-40) ( When $\Lambda = L_G$,

$\frac{\partial(\sqrt{-g}\,\Lambda)}{\partial\psi}=0$, $\frac{\partial(\sqrt{-g}\,\Lambda)}{\partial\psi_{,\lambda}}=0$ ). Eq.(36) means that $\Gamma^{ij}_{..\mu,\nu}(x)$ appear in $\sqrt{-g}\,L_G(x)$ only through curvature tensor field $R^{ij}_{..\mu\nu}(x)$ because $2\frac{\partial(\sqrt{-g}\,L_G)}{\partial R^{ij}_{..\nu\mu}}\equiv\frac{\partial(\sqrt{-g}\,L_G)}{\partial\Gamma^{ij}_{..\mu,\nu}}$ ; Eq.(40) means that $h^{i}_{.\mu,\nu}(x)$ appear in $\sqrt{-g}\,L_G(x)$ only through torsion tensor field $T^{i}_{.\mu\nu}(x)$ because $\frac{\partial(\sqrt{-g}\,L_G)}{\partial T^{i}_{.\mu\nu}}\equiv\frac{\partial(\sqrt{-g}\,L_G)}{\partial h^{i}_{.\mu,\nu}}$ ; Eq.(35) means that $\Gamma^{mn}_{..\nu}(x)$ appear in $\sqrt{-g}\,L_G(x)$ only through curvature tensor field $R^{ij}_{..\mu\nu}(x)$ and torsion tensor field $T^{i}_{.\mu\nu}(x)$ because

$$\frac{\partial(\sqrt{-g}\,L_G)}{\partial\Gamma^{mn}_{..\nu}}=\frac{\partial(\sqrt{-g}\,L_G)}{\partial R^{ij}_{..\alpha\beta}}\frac{\partial R^{ij}_{..\alpha\beta}}{\partial\Gamma^{mn}_{..\nu}}+\frac{\partial(\sqrt{-g}\,L_G)}{\partial T^{i}_{.\alpha\beta}}\frac{\partial T^{i}_{.\alpha\beta}}{\partial\Gamma^{mn}_{..\nu}}$$
$$=2\frac{\partial(\sqrt{-g}\,L_G)}{\partial\Gamma^{jm}_{..\mu,\nu}}\Gamma^{j}_{.n\mu}+\frac{\partial(\sqrt{-g}\,L_G)}{\partial h^{m}_{.\mu,\nu}}h_{.n\mu} \qquad (65)$$

hence the gravitational Lagrangian density $\sqrt{-g}\,L_G(x)$ should be taken the form denoted by Eq.(64).

The form of Eq.(64) exhibits that $\sqrt{-g}\,L_G(x)$ of gravitational system with torsion can be composed of curvature tensor field and torsion field. Because $L^{RT}_G(x)$ is both a coordinate scalar and a frame scalar, the possible terms involved in $L^{RT}_G(x)$ are scalars constructed from $R^{ij}_{..\mu\nu}(x), T^{i}_{.\mu\nu}(x), h^{i}_{.\mu}(x)$.

If Eq.(4) is used to describe a gravitational system without torsion, then $T^{i}_{.\mu\nu}(x)=0$ and $R^{\sigma(\Gamma)}_{.\lambda\mu\nu} = R^{\sigma(\{\})}_{.\lambda\mu\nu}$ ; the possible terms involved in $L_G(x)$ are only the scalar curvature $R^{\{\}} = h_i^{\cdot\mu} h_j^{\cdot j} R^{ij}_{..\mu\nu}$ and its power such as $R^{2\{\}}$ ... . Considering other requirements [4], $L_G(x) = \frac{1}{16\pi G}[R(x) + 2\lambda]$ is chosen in general relativity. We have shown that in Ref.[1], $L_G(x) = \frac{1}{16\pi G}[R(x) + 2\lambda + 2D(x)]$ is another possible choice.

## 5. Conservation laws of energy-momentum tensor density and Conservation laws of spin density derived from the symmetries of Lagrangian density for a gravitational system

It is well known that, in the special relativity, the conservation laws for a physical system is originated from the action integral $I = \int \sqrt{-g(x)} L(x) d^4 x$ of this physical system being invariant under some symmetry of transformations. Similarly in relativistic theories of gravitation, the conservation laws of energy-momentum tensor density and the conservation laws of spin density can be derived from the symmetries denoted by Eq.(22) and Eq.(23) respectively.

Ⅰ **The conservation laws of energy-momentum tensor density**

Let us to start from the conservation laws of energy-momentum tensor density for a gravitational system represented by the generalized Einstein's Lagrangian densities Eqs.(1,2). In Ref.[1], we have derived the Lorentz and Levi-Civita's conservation laws from Eqs.(1,2):

$$\frac{\partial}{\partial x^\lambda}(\sqrt{-g} T^{*\lambda}_{(M)\alpha} + \sqrt{-g} T^{*\lambda}_{(G)\alpha}) = 0 \qquad (66)$$

$$\sqrt{-g} T^{*\lambda}_{(M)\alpha} + \sqrt{-g} T^{*\lambda}_{(G)\alpha} = 0 \qquad (67)$$

Where [1] $\quad \sqrt{-g} T^{*\lambda}_{(M)\alpha} = \sqrt{-g} L_M \delta^\lambda_\alpha - \frac{\partial(\sqrt{-g} L_M)}{\partial \psi_{,\lambda}} \psi_{,\alpha} - \frac{\partial(\sqrt{-g} L_M)}{\partial h^i_{.\mu,\lambda}} h^i_{.\mu,\alpha}$

$$\sqrt{-g}\,T^{*\lambda}_{(G)\alpha} = \sqrt{-g}\,L_G\,\delta^\lambda_\alpha - \frac{\partial(\sqrt{-g}\,L_G)}{\partial h^i_{.\mu,\lambda}}h^i_{.\mu,\alpha} - \frac{\partial(\sqrt{-g}\,L_G)}{\partial h^i_{.\mu,\lambda\sigma}}h^i_{.\mu,\sigma\alpha}$$

$$+ \frac{\partial}{\partial x^\sigma}\left(\frac{\partial(\sqrt{-g}\,L_G)}{\partial h^i_{.\mu,\lambda\sigma}}\right)h^i_{.\mu,\alpha} + \frac{\partial^2}{\partial x^\mu \partial x^\sigma}\left(\frac{\partial(\sqrt{-g}\,L_G)}{\partial h^i_{.\lambda,\mu\sigma}}h^i_{.\alpha}\right)$$

In Ref.[1], there is another definition for the energy-momentum tensor density of pure gravitational field:

$$\sqrt{-g}\,t^{*\lambda}_{(G)\alpha} = \sqrt{-g}\,L_G\,\delta^\lambda_\alpha - \frac{\partial(\sqrt{-g}\,L_G)}{\partial h^i_{.\mu,\lambda}}h^i_{.\mu,\alpha} - \frac{\partial(\sqrt{-g}\,L_G)}{\partial h^i_{.\mu,\lambda\sigma}}h^i_{.\mu,\sigma\alpha} + \frac{\partial}{\partial x^\sigma}\left(\frac{\partial(\sqrt{-g}\,L_G)}{\partial h^i_{.\mu,\lambda\sigma}}\right)h^i_{.\mu,\alpha}$$

and correspondingly there is another conservation laws:

$$\frac{\partial}{\partial x^\lambda}\left(\sqrt{-g}\,T^{*\lambda}_{(M)\alpha} + \sqrt{-g}\,t^{*\lambda}_{(G)\alpha}\right) = 0 \tag{68}$$

which are called the Einstein's conservation laws. We have shown in Ref.[1] that Eq.(68) is equivalent to Eq.(66) in mathematical sense, but they are different in physical sense. Because of the reasons expounded in Ref.[10] and Ref.[1], Eq.(66) might be better than Eq.(68).

In Ref.[1], there has yet the relation:

$$\sqrt{-g}\,T^{*\lambda}_{(M)\alpha} = \sqrt{-g}\,L_M\,\delta^\lambda_\alpha - \frac{\partial(\sqrt{-g}\,L_M)}{\partial \psi_{,\lambda}}\psi_{,\alpha} - \frac{\partial(\sqrt{-g}\,L_M)}{\partial h^i_{.\mu,\lambda}}h^i_{.\mu,\alpha}$$

$$= h^i_{.\alpha}\left(\frac{\partial(\sqrt{-g}\,L_M)}{\partial h^i_{.\lambda}} - \frac{\partial}{\partial x^\mu}\left(\frac{\partial(\sqrt{-g}\,L_M)}{\partial h^i_{.\lambda,\mu}}\right)\right) \tag{69}$$

we shall use Eq.(69) below.

For the more generalized Lagrangian densities Eqs.(3,4), Eqs.(37-39) stem from the symmetry of local space-time translations Eq.(22), and the conservation laws of energy-momentum tensor density for a gravitational system can be derived from these identities. Let $\Lambda = L_M + L_G$, Eq.(37) can be transformed into

$$\frac{\partial}{\partial x^\lambda}(\sqrt{-g}(L_M + L_G)\delta_\alpha^\lambda - \frac{\partial(\sqrt{-g} L_M)}{\partial \psi_{,\lambda}}\psi_{,\alpha} - \frac{\partial(\sqrt{-g}(L_M + L_G))}{\partial h^i_{.\mu,\lambda}}h^i_{.\mu,\alpha}$$

$$-\frac{\partial(\sqrt{-g} L_G)}{\partial \Gamma^{ij}_{..\mu,\lambda}}\Gamma^{ij}_{..\mu,\alpha}) = (\frac{\partial(\sqrt{-g} L_M)}{\partial \psi} - \frac{\partial}{\partial x^\lambda}(\frac{\partial(\sqrt{-g} L_M)}{\partial \psi_{,\lambda}}))\psi_{,\alpha}$$

$$+ (\frac{\partial(\sqrt{-g}(L_M + L_G))}{\partial h^i_{.\mu}} - \frac{\partial}{\partial x^\lambda}(\frac{\partial(\sqrt{-g}(L_M + L_G))}{\partial h^i_{.\mu,\lambda}})) h^i_{.\mu,\alpha}$$

$$+ (\frac{\partial(\sqrt{-g}(L_M + L_G))}{\partial \Gamma^{ij}_{..\mu}} - \frac{\partial}{\partial x^\lambda}(\frac{\partial(\sqrt{-g} L_G)}{\partial \Gamma^{ij}_{..\mu,\lambda}})) \Gamma^{ij}_{..\mu,\alpha} \qquad (70)$$

Utilizing the equations of fields Eqs.(18,19,20), we get from Eq.(70)

$$\frac{\partial}{\partial x^\lambda}(\sqrt{-g}(L_M + L_G)\delta_\alpha^\lambda - \frac{\partial(\sqrt{-g} L_M)}{\partial \psi_{,\lambda}}\psi_{,\alpha} - \frac{\partial(\sqrt{-g}(L_M + L_G))}{\partial h^i_{.\mu,\lambda}}h^i_{.\mu,\alpha} \qquad (71)$$

$$-\frac{\partial(\sqrt{-g} L_G)}{\partial \Gamma^{ij}_{..\mu,\lambda}}\Gamma^{ij}_{..\mu,\alpha}) = 0$$

Eq.(71) might be regarded as conservation laws of energy-momentum tensor density for gravitational system. Because in the special relativity, the expression $L_M \delta_\alpha^\lambda - \frac{\partial L_M}{\partial \psi_{,\lambda}}\psi_{,\alpha}$ is the energy-momentum tensor of matter field, then

$$\sqrt{-g}T'^\lambda_{(M)\alpha} = \sqrt{-g} L_M \delta_\alpha^\lambda - \frac{\partial(\sqrt{-g} L_M)}{\partial \psi_{,\lambda}}\psi_{,\alpha} - \frac{\partial(\sqrt{-g} L_M)}{\partial h^i_{.\mu,\lambda}}h^i_{.\mu,\alpha} \qquad (72)$$

and $\qquad \sqrt{-g}T'^\lambda_{(G)\alpha} = \sqrt{-g} L_G \delta_\alpha^\lambda - \frac{\partial(\sqrt{-g} L_G)}{\partial h^i_{.\mu,\lambda}}h^i_{.\mu,\alpha} - \frac{\partial(\sqrt{-g} L_G)}{\partial \Gamma^{ij}_{..\mu,\lambda}}\Gamma^{ij}_{..\mu,\alpha} \qquad (73)$

might be interpreted as the energy-momentum tensor density of matter field and of gravitational field respectively. But the energy-momentum tensor density of matter field $\sqrt{-g}T^\lambda_{(M)\alpha}$ is defined historically

by [3]

$$\sqrt{-g}\, T^{\lambda}_{(M)\alpha} = h^{i}_{.\alpha}\left(\frac{\partial(\sqrt{-g}\, L_M)}{\partial h^{i}_{.\lambda}} - \frac{\partial}{\partial x^{\mu}}\left(\frac{\partial(\sqrt{-g}\, L_M)}{\partial h^{i}_{.\lambda,\mu}}\right)\right) \tag{74}$$

and this definition remains unchanged in the theories of gravitation till now. We will use this definition in this paper. As shown by Eq.(69), the definition Eq.(74) is identical with Eq.(72) in the theories of gravitation without torsion; but we shall show below that these two definitions are different in the theories of gravitation with torsion. The differences of physical properties caused by these two definitions are well worth to study, which will not be the topic of this paper.

Being similar to Eq.(74), $\sqrt{-g}\, T^{\lambda}_{(G)\alpha}$ is defined by

$$\sqrt{-g}\, T^{\lambda}_{(G)\alpha} = h^{i}_{.\alpha}\left(\frac{\partial(\sqrt{-g}\, L_G)}{\partial h^{i}_{.\lambda}} - \frac{\partial}{\partial x^{\mu}}\left(\frac{\partial(\sqrt{-g}\, L_G)}{\partial h^{i}_{.\lambda,\mu}}\right)\right) \tag{75}$$

From the equations of field Eq.(19), we immediately obtain

$$\sqrt{-g}\, T^{\lambda}_{(M)\alpha} + \sqrt{-g}\, T^{\lambda}_{(G)\alpha} = 0 \tag{76}$$

$$\frac{\partial}{\partial x^{\lambda}}\left(\sqrt{-g}\, T^{\lambda}_{(M)\alpha} + \sqrt{-g}\, T^{\lambda}_{(G)\alpha}\right) = 0 \tag{77}$$

We will call Eqs.(76,77) the Lorentz and Levi-Civita's conservation laws for the gravitational system with the more generalized Lagrangian densities denoted by Eqs.(3,4). The definitions Eqs.(74,75) is not identical with Eqs.(72,73), but it must be indicated that by using the definitions Eqs.(72,73) and from Eqs.(19,20,38,39,71) we can obtain also

$$\sqrt{-g}\, T'^{\lambda}_{(M)\alpha} + \sqrt{-g}\, T'^{\lambda}_{(G)\alpha} = 0$$

$$\frac{\partial}{\partial x^{\lambda}}\left(\sqrt{-g}\, T'^{\lambda}_{(M)\alpha} + \sqrt{-g}\, T'^{\lambda}_{(G)\alpha}\right) = 0$$

Let $\Lambda = L_M + L_G$, Eq.(38) can be transformed into

$$\sqrt{-g}(L_M + L_G)\delta_\alpha^\lambda - \frac{\partial(\sqrt{-g}\,L_M)}{\partial\psi_{,\lambda}}\psi_{,\alpha} - \frac{\partial(\sqrt{-g}(L_M + L_G))}{\partial h^i_{.\mu,\lambda}}h^i_{.\mu,\alpha}$$

$$-\frac{\partial(\sqrt{-g}\,L_G)}{\partial\Gamma^{ij}_{..\mu,\lambda}}\Gamma^{ij}_{..\mu,\alpha} = \left(\frac{\partial(\sqrt{-g}(L_M + L_G))}{\partial h^i_{.\lambda}} - \frac{\partial}{\partial x^\mu}\left(\frac{\partial(\sqrt{-g}(L_M + L_G))}{\partial h^i_{.\lambda,\mu}}\right)\right)h^i_{.\alpha}$$

$$+\left(\frac{\partial(\sqrt{-g}(L_M + L_G))}{\partial\Gamma^{ij}_{..\lambda}} - \frac{\partial}{\partial x^\mu}\left(\frac{\partial(\sqrt{-g}\,L_G)}{\partial\Gamma^{ij}_{..\lambda,\mu}}\right)\right)\Gamma^{ij}_{..\alpha}$$

$$+\frac{\partial}{\partial x^\mu}\left(\left(\frac{\partial(\sqrt{-g}(L_M + L_G))}{\partial h^i_{.\lambda,\mu}}\right)h^i_{.\alpha} + \left(\frac{\partial(\sqrt{-g}\,L_G)}{\partial\Gamma^{ij}_{..\lambda,\mu}}\right)\Gamma^{ij}_{..\alpha}\right) \qquad (78)$$

Let $\Lambda = L_M$, Eq.(78) become

$$\sqrt{-g}\,L_M\,\delta_\alpha^\lambda - \frac{\partial(\sqrt{-g}\,L_M)}{\partial\psi_{,\lambda}}\psi_{,\alpha} - \frac{\partial(\sqrt{-g}\,L_M)}{\partial h^i_{.\mu,\lambda}}h^i_{.\mu,\alpha}$$

$$= \left(\frac{\partial(\sqrt{-g}\,L_M)}{\partial h^i_{.\lambda}} - \frac{\partial}{\partial x^\mu}\left(\frac{\partial(\sqrt{-g}\,L_M)}{\partial h^i_{.\lambda,\mu}}\right)\right)h^i_{.\alpha} + \left(\frac{\partial(\sqrt{-g}\,L_M)}{\partial\Gamma^{ij}_{..\lambda}}\right)\Gamma^{ij}_{..\alpha} + \frac{\partial}{\partial x^\mu}\left(\left(\frac{\partial(\sqrt{-g}\,L_M)}{\partial h^i_{.\lambda,\mu}}\right)h^i_{.\alpha}\right) \qquad (79)$$

If the definition Eq.(72) is adopted, we get from Eq.(79):

$$\sqrt{-g}\,T'^\lambda_{(M)\alpha} = \left(\frac{\partial(\sqrt{-g}\,L_M)}{\partial h^i_{.\lambda}} - \frac{\partial}{\partial x^\mu}\left(\frac{\partial(\sqrt{-g}\,L_M)}{\partial h^i_{.\lambda,\mu}}\right)\right)h^i_{.\alpha}$$

$$+\left(\frac{\partial(\sqrt{-g}\,L_M)}{\partial\Gamma^{ij}_{..\lambda}}\right)\Gamma^{ij}_{..\alpha} + \frac{\partial}{\partial x^\mu}\left(\left(\frac{\partial(\sqrt{-g}\,L_M)}{\partial h^i_{.\lambda,\mu}}\right)h^i_{.\alpha}\right) \qquad (80)$$

If the definition Eq.(74) is adopted, we get from Eq.(79):

$$\sqrt{-g}\,L_M\,\delta_\alpha^\lambda - \frac{\partial(\sqrt{-g}\,L_M)}{\partial\psi_{,\lambda}}\psi_{,\alpha} - \frac{\partial(\sqrt{-g}\,L_M)}{\partial h^i_{.\mu,\lambda}}h^i_{.\mu,\alpha}$$

$$= \sqrt{-g}\,T^\lambda_{(M)\alpha} + \left(\frac{\partial(\sqrt{-g}\,L_M)}{\partial\Gamma^{ij}_{..\lambda}}\right)\Gamma^{ij}_{..\alpha} + \frac{\partial}{\partial x^\mu}\left(\left(\frac{\partial(\sqrt{-g}\,L_M)}{\partial h^i_{.\lambda,\mu}}\right)h^i_{.\alpha}\right) \qquad (81)$$

Evidently $\sqrt{-g}\, T'^{\lambda}_{(G)\alpha}$ is different from $\sqrt{-g}\, T^{\lambda}_{(G)\alpha}$.

Ⅱ **The conservation laws of spin density**

Let $\Lambda = L_M + L_G$, Eq.(34) can be transformed into

$$\frac{\partial}{\partial x^\lambda}\left(\frac{1}{2}\frac{\partial(\sqrt{-g}\,L_M)}{\partial \psi_{,\lambda}}\sigma_{mn}\psi + \frac{\partial(\sqrt{-g}(L_M+L_G))}{\partial h^m_{\cdot\mu,\lambda}}h_{n\mu}\right.$$

$$\left. + 2\frac{\partial(\sqrt{-g}\,L_G)}{\partial \Gamma^{km}_{..\mu,\lambda}}\Gamma^k_{.n\mu}\right) = \frac{1}{2}\left(\frac{\partial(\sqrt{-g}\,L_M)}{\partial \psi} - \frac{\partial}{\partial x^\lambda}\left(\frac{\partial(\sqrt{-g}\,L_M)}{\partial \psi_{,\lambda}}\right)\right)\sigma_{mn}\psi$$

$$+ \left(\frac{\partial(\sqrt{-g}(L_M+L_G))}{\partial h^m_{\cdot\mu}} - \frac{\partial}{\partial x^\lambda}\left(\frac{\partial(\sqrt{-g}(L_M+L_G))}{\partial h^m_{\cdot\mu,\lambda}}\right)\right)h_{n\mu}$$

$$+ 2\left(\frac{\partial(\sqrt{-g}(L_M+L_G))}{\partial \Gamma^{km}_{..\mu}} - \frac{\partial}{\partial x^\lambda}\left(\frac{\partial(\sqrt{-g}\,L_G)}{\partial \Gamma^{km}_{..\mu,\lambda}}\right)\right)\Gamma^k_{.n\mu}$$

(82)

Utilizing the equations of fields Eqs.(18,19,20), we get from Eq.(82)

$$\frac{\partial}{\partial x^\lambda}\left(\frac{1}{2}\frac{\partial(\sqrt{-g}\,L_M)}{\partial \psi_{,\lambda}}\sigma_{mn}\psi + \frac{\partial(\sqrt{-g}(L_M+L_G))}{\partial h^m_{\cdot\mu,\lambda}}h_{n\mu}\right.$$

$$\left. + 2\frac{\partial(\sqrt{-g}\,L_G)}{\partial \Gamma^{km}_{..\mu,\lambda}}\Gamma^k_{.n\mu}\right) = 0$$

(83)

Letting $\Lambda = L_M + L_G$ and utilizing the equations of field Eq.(20), Eq.(35) can be reduced to

$$\frac{1}{2}\frac{\partial(\sqrt{-g}\,L_M)}{\partial \psi_{,\lambda}}\sigma_{mn}\psi + \frac{\partial(\sqrt{-g}(L_M+L_G))}{\partial h^m_{\cdot\mu,\lambda}}h_{n\mu} + 2\frac{\partial(\sqrt{-g}\,L_G)}{\partial \Gamma^{km}_{..\mu,\lambda}}\Gamma^k_{.n\mu}$$

$$- \frac{\partial}{\partial x^\mu}\left(\frac{\partial(\sqrt{-g}\,L_G)}{\partial \Gamma^{mn}_{..\lambda,\mu}}\right) = \frac{\partial(\sqrt{-g}(L_M+L_G))}{\partial \Gamma^{mn}_{..\lambda}} - \frac{\partial}{\partial x^\mu}\left(\frac{\partial(\sqrt{-g}\,L_G)}{\partial \Gamma^{mn}_{..\lambda,\mu}}\right) = 0$$

(84)

Eq.(83) can be obtained from Eq.(84) by differentiation.

For the gravitational system with torsion, we define the spin tensor of matter field and the spin tensor of gravitational field as

$$\sqrt{-g}\, S^{\lambda}_{(M)mn} = \frac{\partial(\sqrt{-g}\, L_M)}{\partial \Gamma^{mn}_{..\lambda}} \tag{85}$$

and
$$\sqrt{-g}\, S^{\lambda}_{(G)mn} = \frac{\partial(\sqrt{-g}\, L_G)}{\partial \Gamma^{mn}_{..\lambda}} - \frac{\partial}{\partial x^{\mu}}\left(\frac{\partial(\sqrt{-g}\, L_G)}{\partial \Gamma^{mn}_{..\lambda,\mu}}\right) \tag{86}$$

respectively; which are similar to the definitions of Kibble [3] except with different multiples and signs. Then from Eq.(84) and Eq.(83) we get

$$\sqrt{-g}\, S^{\lambda}_{(M)mn} = \frac{1}{2}\frac{\partial(\sqrt{-g}\, L_M)}{\partial \psi_{,\lambda}}\sigma_{mn}\psi + \frac{\partial(\sqrt{-g}\, L_M)}{\partial h^{m}_{.\mu,\lambda}} h_{n\mu} \tag{87}$$

$$\sqrt{-g}\, S^{\lambda}_{(G)mn} = \frac{\partial(\sqrt{-g}\, L_G)}{\partial h^{m}_{.\mu,\lambda}} h_{n\mu} + 2\frac{\partial(\sqrt{-g}\, L_G)}{\partial \Gamma^{km}_{..\mu,\lambda}}\Gamma^{k}_{.n\mu} - \frac{\partial}{\partial x^{\mu}}\left(\frac{\partial(\sqrt{-g}\, L_G)}{\partial \Gamma^{mn}_{..\lambda,\mu}}\right) \tag{88}$$

$$\frac{\partial}{\partial x^{\lambda}}\left(\sqrt{-g}\, S^{\lambda}_{(M)mn} + \sqrt{-g}\, S^{\lambda}_{(G)mn}\right) = 0 \tag{89}$$

and
$$\sqrt{-g}\, S^{\lambda}_{(M)mn} + \sqrt{-g}\, S^{\lambda}_{(G)mn} = 0 \tag{90}$$

Eqs.(89,90) are the conservation laws of spin density for gravitational system denoted by the more generalized Lagrangian densities Eqs.(3,4). In the special relativity, $\frac{1}{2}\frac{\partial L_M}{\partial \psi_{,\lambda}}\sigma_{mn}\psi$ is the spin tensor of matter field, $\frac{\partial(\sqrt{-g}\, L_M)}{\partial h^{m}_{.\mu,\lambda}} h_{n\mu}$ should be interpreted as the influence of gravitational field.

Letting $\Lambda = L_M$ and utilizing Eqs.(18,75,85,87), Eq.(82) can be reduced to

$$\frac{\partial}{\partial x^{\lambda}}\left(\sqrt{-g}\, S^{\lambda}_{(M)mn}\right) = \sqrt{-g}\, T_{(M)[mn]} + \sqrt{-g}\, \Gamma^{k}_{.m\lambda} S^{\lambda}_{(M)kn} + \sqrt{-g}\, \Gamma^{k}_{.n\lambda} S^{\lambda}_{(M)mk} \tag{91}$$

This is an important equation for studying the dynamical problems involving spin.

For the generalized Einstein's Lagrangian densities Eqs.(1,2), the conservation laws of spin density can be derived from Eqs.(44-46) with the similar methods. Let $\Lambda = L_M + L_G$, Eq.(44) can be transformed into

$$\frac{\partial}{\partial x^\lambda}(\frac{1}{2}\frac{\partial(\sqrt{-g}\,L_M)}{\partial \psi_{,\lambda}}\sigma_{mn}\psi + \frac{\partial(\sqrt{-g}\,(L_M+L_G))}{\partial h^m_{\cdot\mu,\lambda}}h_{n\mu}$$
$$+ \frac{\partial(\sqrt{-g}\,L_G)}{\partial h^m_{\cdot\mu,\lambda\sigma}}h_{n\mu,\sigma} - [\frac{\partial}{\partial x^\sigma}(\frac{\partial(\sqrt{-g}\,L_G)}{\partial h^m_{\cdot\mu,\lambda\sigma}})]h_{n\mu})$$
$$= \frac{1}{2}(\frac{\partial(\sqrt{-g}\,L_M)}{\partial \psi} - \frac{\partial}{\partial x^\lambda}(\frac{\partial(\sqrt{-g}\,L_M)}{\partial \psi_{,\lambda}}))\sigma_{mn}\psi$$
$$+ (\frac{\partial(\sqrt{-g}\,(L_M+L_G))}{\partial h^m_{\cdot\mu}} - \frac{\partial}{\partial x^\lambda}(\frac{\partial(\sqrt{-g}\,(L_M+L_G))}{\partial h^m_{\cdot\mu,\lambda}}) + \frac{\partial^2}{\partial x^\lambda \partial x^\sigma}(\frac{\partial(\sqrt{-g}\,L_G)}{\partial h^m_{\cdot\mu,\lambda\sigma}}))h_{n\mu}$$

(92)

Utilizing the equations of fields Eqs.(18',21), we get from Eq.(92)

$$\frac{\partial}{\partial x^\lambda}(\frac{1}{2}\frac{\partial(\sqrt{-g}\,L_M)}{\partial \psi_{,\lambda}}\sigma_{mn}\psi + \frac{\partial(\sqrt{-g}\,(L_M+L_G))}{\partial h^m_{\cdot\mu,\lambda}}h_{n\mu}$$
$$+ \frac{\partial(\sqrt{-g}\,L_G)}{\partial h^m_{\cdot\mu,\lambda\sigma}}h_{n\mu,\sigma} - [\frac{\partial}{\partial x^\sigma}(\frac{\partial(\sqrt{-g}\,L_G)}{\partial h^m_{\cdot\mu,\lambda\sigma}})]h_{n\mu}) = 0$$

(93)

If we define

$$\sqrt{-g}\,S^{*\lambda}_{(M)mn} = \frac{1}{2}\frac{\partial(\sqrt{-g}\,L_M)}{\partial \psi_{,\lambda}}\sigma_{mn}\psi + \frac{\partial(\sqrt{-g}\,L_M)}{\partial h^m_{\cdot\mu,\lambda}}h_{n\mu}$$

(94)

$$\sqrt{-g}\,S^{*\lambda}_{(G)mn} = \frac{\partial(\sqrt{-g}\,L_G)}{\partial h^m_{\cdot\mu,\lambda}}h_{n\mu} + \frac{\partial(\sqrt{-g}\,L_G)}{\partial h^m_{\cdot\mu,\lambda\sigma}}h_{n\mu,\sigma} - [\frac{\partial}{\partial x^\sigma}(\frac{\partial(\sqrt{-g}\,L_G)}{\partial h^m_{\cdot\mu,\lambda\sigma}})]h_{n\mu}$$

(95)

Eq.(93) becomes

$$\frac{\partial}{\partial x^\lambda}(\sqrt{-g}\, S^{*\lambda}_{(M)mn} + \sqrt{-g}\, S^{*\lambda}_{(G)mn}) = 0 \tag{96}$$

Letting $\Lambda = L_M + L_G$ and utilizing the identity Eq.(46), and noting

$$\varepsilon^{mn}_{,\lambda}\frac{\partial}{\partial x^\sigma}(\frac{\partial L_G}{\partial h^m_{.\mu,\lambda\sigma}} h_{n\mu}) = 0, \text{ Eq.(45) can be transformed into}$$

$$\sqrt{-g}\, S^{*\lambda}_{(M)mn} + \sqrt{-g}\, S^{*\lambda}_{(G)mn} = 0 \tag{97}$$

Letting $\Lambda = L_M$ and utilizing Eqs.(18',94) and using the relation [1]

$$\sqrt{-g}\, T^{*\lambda}_{(M)\alpha} \overset{def}{=} \sqrt{-g}\, L_M \delta^\lambda_\alpha - \frac{\partial(\sqrt{-g}\, L_M)}{\partial \psi_{,\lambda}} \psi_{,\alpha} - \frac{\partial(\sqrt{-g}\, L_M)}{\partial h^i_{.\mu,\lambda}} h^i_{.\mu,\alpha}$$

$$= h^i_{.\alpha}(\frac{\partial(\sqrt{-g}\, L_M)}{\partial h^i_{.\lambda}} - \frac{\partial}{\partial x^\mu}(\frac{\partial(\sqrt{-g}\, L_M)}{\partial h^i_{.\lambda,\mu}}))$$

Eq.(92) can be reduced to

$$\frac{\partial}{\partial x^\lambda}(\sqrt{-g}\, S^{*\lambda}_{(M)mn}) = \sqrt{-g}\, T^*_{(M)[mn]} \tag{98}$$

The appearance of $T_{[mn]}$ in the above equation is due to $S^\lambda_{(M)mn} = -S^\lambda_{(M)nm}$.

Since the conservation laws of spin density for a gravitational system require studying specially, we will not go a step further to discuss the specific properties about these conservation laws of spin density in this paper.

## 6. Energy-momentum equations of motion and spin equations of motion for test particle

The so-called test particle is a very small material body used to test the gravitational actions; it is assumed that both the space extension of the test particle and its self-gravitation are extremely small and can be neglected in the discussed problems. The test particle possesses energy-momentum and spin [5]; its energy-momentum vector $p_\nu$ and spin angular momentum $S_{ij}$ can be determined by

$$p_\nu = \int \sqrt{-g}\, T^0_{(M)\nu}\, d^3 x \tag{99}$$

and
$$S_{ij} = \int \sqrt{-g}\, S^0_{(M)ij}\, d^3x \qquad (100)$$

respectively. There are two equations of motion for test particle, *i.e.* energy-momentum equations of motion and spin equations of motion. These two equations of motion will be derived below from the conservation laws of energy-momentum tensor density and the conservation laws of spin density for a gravitational system respectively.

Ⅰ **The energy-momentum equations of motion**

For the gravitational system with torsion which can be denoted by the more generalized Lagrangian densities Eqs.(3,4); Letting $\Lambda = L_M$ and utilizing Eqs.(18,74,85), Eq.(70) can be reduced to

$$\frac{\partial}{\partial x^\lambda}\left(\sqrt{-g}\, L_M\, \delta^\lambda_\alpha - \frac{\partial(\sqrt{-g}\, L_M)}{\partial \psi_{,\lambda}}\psi_{,\alpha} - \frac{\partial(\sqrt{-g}\, L_M)}{\partial h^i_{\cdot\mu,\lambda}} h^i_{\cdot\mu,\alpha}\right) \qquad (101)$$
$$= \sqrt{-g}\, T^{\cdot\mu}_{(M)i}\, h^i_{\cdot\mu,\alpha} + \sqrt{-g}\, S^{\cdot\cdot\mu}_{(M)ij}\, \Gamma^{ij}_{\cdot\cdot\mu,\alpha}$$

From Eq.(81) and Eq.(101) and utilizing Eq.(91), we can get the "conservation laws" of the energy-momentum tensor density of matter field [5]:

$$\frac{\partial}{\partial x^\mu}(\sqrt{-g}\, T^{\cdot\mu}_{(M)\nu}) - \Gamma^\sigma_{\cdot\mu\nu}\sqrt{-g}\, T^{\cdot\mu}_{(M)\sigma} + \sqrt{-g}\, S^{\cdot\cdot\mu}_{(M)ij}\, R^{ij}_{\cdot\cdot\mu\nu} = 0 \qquad (102)$$

For a test particle, by using the method of Papapetrou [11], the energy-momentum equation of motion

$$\frac{dp_\nu}{dt} - \Gamma^\sigma_{\mu\nu}\, p_\sigma\, \frac{dx^\mu}{dt} + R^{ij}_{\cdot\cdot\mu\nu}\, S_{ij}\, \frac{dx^\mu}{dt} = 0 \qquad (103)$$

can be obtained from Eqs.(99,100).

For the gravitational system without torsion which can be denoted by the generalized Einstein's Lagrangian densities Eqs.(1,2); there exists the relations [1] :

$$\sqrt{-g}\, T^{*\lambda}_{(M)\alpha} = \sqrt{-g}\, L_M\, \delta^\lambda_\alpha - \frac{\partial(\sqrt{-g}\, L_M)}{\partial \psi_{,\lambda}}\psi_{,\alpha} - \frac{\partial(\sqrt{-g}\, L_M)}{\partial h^i_{\cdot\mu,\lambda}} h^i_{\cdot\mu,\alpha} \qquad (104)$$

$$\frac{\partial}{\partial x^\lambda}(\sqrt{-g}\, L_M\, \delta^\lambda_\alpha - \frac{\partial(\sqrt{-g}\, L_M)}{\partial \psi_{,\lambda}}\psi_{,\alpha} - \frac{\partial(\sqrt{-g}\, L_M)}{\partial h^i_{.\mu,\lambda}}h^i_{.\mu,\alpha}) \qquad (105)$$

$$= (\frac{\partial(\sqrt{-g}\, L_M)}{\partial h^i_{.\mu}} - \frac{\partial}{\partial x^\lambda}(\frac{\partial(\sqrt{-g}\, L_M)}{\partial h^i_{.\mu,\lambda}}))\, h^i_{.\mu,\alpha} = \sqrt{-g}\, T^{*\cdot\mu}_{(M)i}\, h^i_{.\mu,\alpha}$$

Hence there are the following equations:

$$\frac{\partial}{\partial x^\lambda}(\sqrt{-g}\, T^{*\cdot\lambda}_{(M)\alpha}) = \sqrt{-g}\, T^{*\cdot\lambda}_{(M)i}\, h^i_{.\lambda,\alpha}$$

from which the "conservation laws" of the energy-momentum tensor density of matter field in the case without torsion:

$$\frac{\partial}{\partial x^\mu}(\sqrt{-g}\, T^{\cdot\mu}_{(M)\nu}) - \Gamma^\sigma_{.\mu\nu}\sqrt{-g}\, T^{\cdot\mu}_{(M)\sigma} = 0$$

can be obtained. After using the method of Papapetrou [11], we can get the momentum equation of motion in the case without torsion:

$$\frac{dp_\nu}{dt} - \Gamma^\sigma_{\mu\nu}\, p_\sigma\, \frac{dx^\mu}{dt} = 0 \qquad (106)$$

It must be emphasized that Eq.(103) does not reduce to Eq.(106) when torsion is equal to zero; as a matter of fact, in that case, Eq.(103) reduces to $\dfrac{dp_\nu}{dt} - \Gamma^\sigma_{\mu\nu}\, p_\sigma\, \dfrac{dx^\mu}{dt} + R^{\{\}\,ij}_{..\mu\nu} S_{ij}\, \dfrac{dx^\mu}{dt} = 0$. We have pointed out that the gravitational field equations with torsion can not be reduced to the gravitational field equations without torsion; since the gravitational field equations are used in the derivations of Eqs.(103,106), hence Eq.(103) does not reduce to Eq.(106).

Ⅱ The **spin equations of motion**
From Eq.(91):

$$\frac{\partial}{\partial x^\lambda}(\sqrt{-g}\, S^\lambda_{(M)mn}) = \sqrt{-g}\, T^{\cdot\mu}_{(M)[mn]} + \sqrt{-g}\, \Gamma^k_{.m\lambda}\, S^\lambda_{(M)kn} + \sqrt{-g}\, \Gamma^k_{.n\lambda}\, S^\lambda_{(M)mk}$$

by using the method of Papapetrou, the spin equation of motion

$$\frac{dS_{mn}}{dt} = h^\nu_{[m}\, h_{n]\mu}\, p_\nu\, \frac{dx^\mu}{dt} + (\Gamma^k_{.m\mu}\, S_{kn} + \Gamma^k_{.n\mu}\, S_{mk})\frac{dx^\mu}{dt} \qquad (107)$$

can be obtained from Eq.(91).
From Eq.(98):

$$\frac{\partial}{\partial x^\lambda}(\sqrt{-g}\, S^{*\lambda}_{(M)mn}) = \sqrt{-g}\, T^*_{(M)[mn]}$$

By using the method of Papapetrou, the spin equation of motion

$$\frac{dS^*_{mn}}{dt} = 0 \tag{108}$$

can be obtained from Eq.(98).

## 7. Conclusions

The following important conclusions can be summarized from the studies and discussions described in this paper:

(1), The dynamical laws of a gravitational system can be derived from the Lagrangian densities of this system; the Lagrangian densities contain rich information concerning many properties of a physical system including gravitation.

(2), The generalized Einstein's Lagrangian densities Eqs.(1,2) are used only to describe a gravitational system without torsion; but the more generalized Lagrangian densities Eqs.(3,4) can be used to describe a system with or without torsion. If Eqs.(3,4) are used to describe a system with torsion, $h^i_{\cdot\mu}(x)$ and $\Gamma^{ij}_{\cdot\cdot\mu}(x)$ are all the independent gravitational field variables ; if Eqs.(3,4) are used to describe a system without torsion, only $h^i_{\cdot\mu}(x)$ is the independent gravitational field variables.

The gravitational field equations are deduced with respect to the independent gravitational field variables; since the independent gravitational field variables are different according to the system being with or without torsion, the gravitational field equations with torsion can not be reduced to the gravitational field equations without torsion. Moreover, for some dynamical equations, if gravitational field equations are used in their derivations the dynamical equations with torsion also can not be reduced to the dynamical equations without torsion.

(3), We have derived from both Eqs.(1,2) and Eqs.(3,4) the following conservation laws:

$$\frac{\partial}{\partial x^\lambda}(\sqrt{-g}\, T^{*\lambda}_{(M)\alpha} + \sqrt{-g}\, T^{*\lambda}_{(G)\alpha}) = 0 \tag{66}$$

$$\sqrt{-g}\, T^{*\lambda}_{(M)\alpha} + \sqrt{-g}\, T^{*\lambda}_{(G)\alpha} = 0 \tag{67}$$

$$\frac{\partial}{\partial x^\lambda}(\sqrt{-g}\, S^{*\lambda}_{(M)mn} + \sqrt{-g}\, S^{*\lambda}_{(G)mn}) = 0 \tag{96}$$

$$\sqrt{-g}\, S^{*\lambda}_{\cdot(M)mn} + \sqrt{-g}\, S^{*\lambda}_{\cdot(G)mn} = 0 \tag{97}$$

$$\frac{\partial}{\partial x^\lambda}\left(\sqrt{-g}\, T^{\cdot\lambda}_{(M)\alpha} + \sqrt{-g}\, T^{\cdot\lambda}_{(G)\alpha}\right) = 0 \tag{77}$$

$$\sqrt{-g}\, T^{\cdot\lambda}_{(M)\alpha} + \sqrt{-g}\, T^{\cdot\lambda}_{(G)\alpha} = 0 \tag{76}$$

$$\frac{\partial}{\partial x^\lambda}\left(\sqrt{-g}\, S^{\cdot\lambda}_{(M)mn} + \sqrt{-g}\, S^{\cdot\lambda}_{(G)mn}\right) = 0 \tag{89}$$

and
$$\sqrt{-g}\, S^{\cdot\lambda}_{(M)mn} + \sqrt{-g}\, S^{\cdot\lambda}_{(G)mn} = 0 \tag{90}$$

These conservation laws have the same mathematical form. It has suggested in Ref.[1] that one important consequence of Eq.(67) or Eq.(76) would be that the energy of matter field might originate from the gravitational field . Similarly it could be suggested that one important consequence of Eq.(97) or Eq.(90) would be that the spin of matter field might originate from the gravitational field . These suggestions could be tested via future experiments and observations.

(4), There are many different definitions for the energy-momentum tensor density and the spin density of a gravitational system in theories of gravitation, the differences of physical properties caused by these different definitions are well worth for further study.

**References**

1. Chen F. P. 2007, "Field equations and conservation laws derived from the generalized Einstein's Lagrangian density for a gravitational system and their implications to cosmology." Accepted by International Journal of Theoretical Physics.

2. Chen F. P. 1993, "The generalized Lagrangians of gravitational theory with torsion and their invariance under ($\varepsilon^{mn}, \xi^\mu$) transformations." *Science In China* (Series A) **36,** 1226.

3. Kibble T.W.B. 1961, "Lorentz Invariance and the Gravitational Field" *J. Math. Phys.* **2,** 212 .

4. Weinberg S. 1972, "Gravitation and Cosmology", Wiley, New York.

5. Chen F. P. 1990, "General equations of motion for test particles in space-time with torsion." *Inter. J. Theor.*

*Phys.* **29,** 161.

6. Hehl F. W., von der Heyde P. and Kerlick G. D. 1976, "General relativity with spin and torsion: foundations and prospects." *Rev. Mod. Phys.* **48,** 393.

7. Schouten J. A. 1954, "Ricci Calculus", Springer, Berlin.

8. Landau L. and Lifshitz E. 1975, "The Classical Theory of Fields", Translated by M. Hamermesh, Pergamon Press, Oxford.

9. Corson E. M. 1953, "Introduction to Tensors, Spinors and Relativistic Wave Equations",Blackie & Son, London.

10. Chen F. P. 2002, "The restudy on the debate between Einstein and Levi-Civita and the experimental tests." *Spacetime & Substance.* **3,** 161.

11. Papapetrou A. 1951, "Spinning test-particles in general relativity. Ⅰ" *Proceedings of the Royal society of London A.***209,** 248